# RODEO: Robust DE-aliasing autoencOder for Real-time Medical Image Reconstruction

Janki Mehta and Angshul Majumdar

Indraprastha Institute of Information Technology, New Delhi, India

## Abstract -

In this work we address the problem of real-time dynamic medical (MRI and X-Ray CT) image reconstruction from parsimonious samples (Fourier frequency space for MRI and sinogram / tomographic projections for CT). Today the de facto standard for such reconstruction is compressed sensing (CS). CS produces high quality images (with minimal perceptual loss); but such reconstructions are time consuming, requiring solving a complex optimization problem. In this work we propose to 'learn' the reconstruction from training samples using an autoencoder. Our work is based on the universal function approximation capacity of neural networks. The training time for the autoencoder is large, but is offline and hence does not affect performance during operation. During testing / operation, our method requires only a few matrix vector products and hence is significantly faster than CS based methods. In fact, it is fast enough for real-time reconstruction (the images are reconstructed as fast as they are acquired) with only slight degradation of image quality. However, in order to make the autoencoder suitable for our problem, we depart from the standard Euclidean norm cost function of autoencoders and use a robust $l_1$-norm instead. The ensuing problem is solved using the Split Bregman method.

## 1. Introduction

Magnetic Resonance Imaging (MRI), X-Ray Computed Tomography (CT) and Ultrasonography (USG) are the most common medical imaging modalities. MRI is a versatile modality that can produce high quality images; but has a relatively prolonged data acquisition time. CT produces decent quality images and is very fast, but owing to its ionizing radiation, poses to be a health hazard for the patient. USG is fast and safe, but yields low quality images which are not readily decipherable; reading USG requires significant training. Therefore in almost all critical situations, MRI and CT are the primary modes of diagnostic imaging. In this work, we concentrate on reconstruction of MRI and CT images.

In MRI, the data collected from the scanner is in the Fourier space. In MRI lingo this is called the K-space. When the K-space is fully sampled on a uniform Cartesian grid, reconstructing the image is trivial – application of an inverse FFT. Unfortunately, sampling the full K-space is time consuming. Reducing the prolonged data acquisition time had been one of the biggest challenges in MRI since its inception. The only way to reduce the scan time (assuming a single receive coil) is to partially sample the K-space and somehow reconstruct the image using smart recovery technique. In the recent past, Compressed Sensing (CS) has been hugely successful in addressing this problem.

In X-Ray CT, data acquisition is fast. To get a good image quality, the number of tomographic projections need to be very high. However it is well known that being subjected to X-Ray ionizing radiation is a health hazard. Even though CT operates within the 'safety limits' imposed by various health monitoring agencies, increasingly it is getting apparent that CT indeed has harmful effects on the patient's health. A recent academic report [1] indicates that about 1.5%-2% of all cancers in the United States can be directly attributed to CT radiation. There are two ways to reduce exposure. The first is reduce the strength of the X-rays; unfortunately it yields very noisy images. The other way to reduce patient's exposure is to have fewer (but of requisite strength) tomographic projections. The challenge is to reconstruct a good quality images from these few tomographic projections. In CT, the motivation for fewer samples (K-space for MRI and projections for CT) is not the same as MRI; but the underlying problem remains the same – that of reconstructing image from fewer projections. Given the similarity of the problem (with MRI), it is apparent that CS has been successful in this domain as well.

CS is a non-linear inversion operation. It is slow, requiring solution of a non-smooth optimization problem that needs to be solved iteratively. This is not an issue for most diagnostic imaging scenarios;

since the scans are almost always read by a doctor much later than it has been acquired. Therefore the latency between acquisition and reconstruction is of no importance. However there may be situations where we would like to recover the image on-the-fly, i.e. in real-time, e.g. in image-guided surgery or tracking problems (catheter tracking). In such cases, CS based techniques cannot be used. The only option so far had been to collect a large number of samples (K-space or projections) and apply analytic reconstruction methods (Fourier inverse for MRI and Filtered Back Projection for CT) for fast recovery. In this work we take a completely new approach to MRI reconstruction based on the machine learning framework - we will LEARN the non-linear reconstruction operation. It is known that neural networks can act as universal function approximators [2-5]; the first work on this topic was from Kolmogorov [2], later this approximation capacity was proven by Kukrova [3], Hornik [4] and Cybenko [5]. Based on this property, in this work, we propose to 'learn' a CS type non-linear inversion operation using autoencoder. The training phase will be time consuming – but it is offline. The operational phase will be superfast; unlike CS we only need a few matrix-vector multiplications for recovery. We will show that, both for MRI and CT, the proposed method yields decent image quality but is a few orders of magnitude faster than existing techniques.

## 2. Literature Review

Some of the readers may be familiar with compressed sensing (CS) based medical image reconstruction techniques, but for the benefit of the larger readership, we briefly review this topic. First we will discuss about CS based MRI reconstruction and then about CT reconstruction. We will also briefly review autoencoders after discussing medical image reconstruction formulations.

### 2.1. MRI Reconstruction

In MRI, the data is acquired in the K-space. This can be formally represented as follows,

$$y = Fx \qquad (1)$$

where x is the underlying image, F is the Fourier transform and y is the K-space data.

If the K-space is fully sampled, recovering x is trivial. Unfortunately full K-space sampling, this is time consuming. In order to accelerate the scan, the K-space is partially sampled. This is represented as,

$$y = RFx \tag{2}$$

Here R is a sampling operator.

The problem (2) is an under-determined inverse problem with infinitely many solutions. In order to get a plausible solution, one needs to make some assumption about x. CS assumes that the image is sparse in certain transform domains (wavelet / DCT). In most cases the sparsifying transform is orthogonal or tight-frame[1], so (2) can be represented as,

$$y = RFW\alpha \tag{3}$$

Here W is the sparsifying transform and α the sparse coefficients.

CS recovers α by solving an $l_1$-norm minimization problem. This is expressed as,

$$\min_{\alpha} \|\alpha\|_1 \text{ subject to } y = RFW\alpha \tag{4}$$

Once the sparse coefficients are required, getting the image is trivial, one just needs to apply the transform on the recovered coefficients. Ideally one would like to solve an $l_0$-norm minimization problem. But that is known to be NP hard [6]. Hence the relaxation to its closest convex surrogate. The optimality of $l_1$-norm minimization has been proven in CS literature [7, 8]. In practice, (4) is further relaxed and an unconstrained version is solved instead.

$$\min_{\alpha} \|y - RFW\alpha\|_2^2 + \lambda \|\alpha\|_1 \tag{5}$$

This direct formulation for MRI reconstruction via CS was proposed in [9]. Since then variations of the basic formulation has been proposed by others [10, 11].

So far we have discussed about, static MRI reconstruction. The problem in dynamic MRI is somewhat different. Here, the K-space of each frame is partially sampled, so the data acquisition can be expressed as,

---

[1] $Orthogonal: WW^T = I = W^TW$
$Tight-frame: WW^T = I \neq W^TW$

$$y_t = RFx_t, t=1...T \tag{6}$$

where t denotes the frame.

This can be comprehensively represented as,

$$Y = RFX, \ Y = [y_1|...|y_T] \text{ and } X = [x_1|...|x_T] \tag{7}$$

The dynamic MRI sequence X is spatio-temporally correlated. As before, each frame (columns of X) can be sparsely represented in a transform domain like wavelet; this accounts for spatial redundancy. The variation in the temporal direction is smooth. Thus it can be compactly represented in the Fourier domain. Thus spatio-temporal redundancy is essentially whitened as,

$$X = WZF_{1D} \Rightarrow vec(X) = F_{1D}^T \otimes W \ vec(Z) \tag{8}$$

Incorporating the Kronecker formulation into (7) leads to,

$$vec(Y) = (I \otimes RF)(F_{1D}^T \otimes W) \ vec(Z) \tag{9}$$

In simpler terms this can be expressed as,

$$y = \Phi\Psi z \tag{10}$$

where $y = vec(Y), z = vec(Z), \Phi = (I \otimes RF)$ and $\Psi = (F_{1D}^T \otimes W)$.

In this form (10), the problem boils down to our well known CS formulation. The solution z can be obtained by solving the $l_1$-minimization problem. This formulation was proposed by [12, 13]. There are many extensions and improvements of the basic formulation [14-18].

There is yet another class of methods that reconstruct the dynamic MRI sequence as a rank deficient matrix. Consider the formulation (7). In [19, 20] it is argued that the matrix X is rank deficient, since it can be modeled as a linear combination of very few temporal basis functions. Based on this assumption, they proposed solving the inverse problem by matrix factorization,

$$\min_{U,V} \|Y - RF(UV)\|_F^2 \tag{11}$$

where X=UV such that X is low-rank.

Solving the inverse problem via matrix factorization was an academic exercise. It did not yield the same level of accuracy as CS based techniques. However, later studies proposed combining the two [21-24]. In [23, 24] it was shown that good results are obtained when the rank deficiency of the signal in *x-t* space is combined with the sparsity in the *x-f* space. The following optimization problem was proposed to solve for the dynamic MRI sequence

$$\min_X \|Y - RFX\|_F^2 + \lambda_1 \|I \otimes F_{1D} vec(X)\|_1 + \lambda_2 \|X\|_* \tag{12}$$

In recent times, there are a few techniques that proposed learning the sparsifying basis in an adaptive fashion during signal reconstruction. This is dubbed as Blind Compressed Sensing (BCS) [25]. Here it is assumed that the dynamic MRI sequence can be sparsely represented in a learned basis by exploiting the temporal correlation, i.e. $X = ZD$ where $D$ is the sparsifying dictionary and $Z$ is the coefficients. Unlike a fixed sparsifying basis like Fourier, $D$ is learnt simultaneously with $Z$. The corresponding problem is formulated as:

$$\min_{D,Z} \|Y - F(ZD)\|_F^2 + \lambda_1 \|Z\|_1 + \lambda_2 \|D\|_F^2 \tag{13}$$

This formulation was proposed in [26]. An extension of this basic formulation was proposed in [27].

So far we have discussed offline retrospective reconstruction techniques. A straightforward way to handle the online dynamic MR reconstruction is to use Kalman Filtering. That is exactly what is done by [28, 29] MRI reconstruction is formulated as a dynamical system and a Kalman filter model is proposed,

$$x_t = x_{t-1} + u_t \tag{14a}$$

$$y_t = RFx_t + \eta_t \tag{14b}$$

where the pixel values $x_t$ is the state variable, the K-space sample $y_t$ is the observation, $\eta_t$ is the observation noise and $u_t$ is the innovation in state variables.

In general the Kalman filter is computationally intensive since it requires explicit matrix inversion for computing the covariance matrix. However [28, 29] alleviated the computational issue by diagonalization of the covariance matrix. This diagonalization was only possible under some simplifying assumptions.

The problem with such a diagonality assumption is that the pixel values are assumed to change independently over time; in other words every point is changing independently of each other. This is not realistic – defies the laws of semi-rigid body dynamics. The diagonality assumption was also subject to sampling requirements – the sampling had to be non-uniform and non-Cartesian. In most practical systems, the sampling is Cartesian.

The issues in the previous work was partially addressed by [30]; Kalman filtering of the wavelet coefficients was proposed therein. The filter model is the following,

$$\alpha_t = \alpha_{t-1} + u_t \tag{15a}$$

$$y_t = RFW\alpha_t + \eta_t \tag{15b}$$

This work was motivated by the findings of compressed sensing. The benefit of this approach is that wavelet whitens the spatial correlations, and hence the diagonality assumption is more feasible compared to the previous studies that operated in the pixel domain. In [30] a further check over the Kalman update was imposed – when the error between the prediction and the actual data becomes large, instead of a filter update full reconstruction using CS technique is done.

A compressed sensing based solution was proposed in [31, 32] for the online dynamic MRI reconstruction problem. It was postulated that the difference between a reference frame and the actual frame should be sparse. In [30], the reference frame was taken as the previous frame; in [31] the reference frame was predicted using an auto-regressive filter. The sparse difference frame can be directly estimated using CS techniques and added to the previous frame.

$$\min_{\nabla x_t} \left\| y_t - RF(x_{ref} + \nabla x_t) \right\|_2^2 + \lambda \left\| \nabla x_t \right\|_1 \tag{16}$$

where $x_t = x_{ref} + \nabla x_t$; ref indicates reference frame.

This simple technique (16) yields better results than all previous Kalman Filtering based methods. We have given a brief survey on this topic; only portions pertinent to our research. For the interested reader we suggest perusing the books [33, 34].

## 2.2. CT Reconstruction

In CT, the data acquisition model can be expressed as follows:

$$y = Ax \qquad (17)$$

Here $x$ is the underlying image (to be reconstructed), $y$ is the sampled sinogram and A is the X-ray transform. For the 2D case using parallel beams, A is just a radon transform; but for more practical scenarios using fan beams, the X-ray transform is more complicated.

For dynamic CT, the sinogram is sampled in an interleaved fashion, so the A changes with time. The data acquisition model for the $t^{th}$ frame is as follows:

$$y_t = A_t x_t \qquad (18)$$

This is an inverse problem; one is supposed to reconstruct $x_t$ given $A_t$ and $y_t$. For non-iterative reconstruction using FBP, infinite number of samples from the sinogram is required, i.e. the angle between two successive tomographic projections should be infinitesimally small. In practice it is not possible to adhere to this requirement; but a large number of samples need to be collected nevertheless. The requirement for large number of projections / samples implies that the subject has to be subjected to more harmful ionizing radiation. Researchers in CT reconstruction want to reconstruct the image from smaller number of sinogram samples.

CS based techniques have been able to achieve this goal. The CT images can be modeled as i) piecewise linear functions, or ii) smooth functions with finite number of discontinuities. The first model leads to a sparse representation of the images in gradient transform (finite differencing) while the second model leads to a sparse representation in wavelet transform. CS exploits the sparsity of the image in order to reconstruct it from lower number of sinogram samples than was deemed necessary previously [35-37]. This technique is suitable for reconstructing static CT images.

In principle one can directly use the techniques developed for static CT and apply it for frame-by-frame reconstruction of dynamic CT sequences. Recent papers have shown that further improvements are possible [38-40]. From the interleaved projections, they generate a static FBP reference image ($x_0$). Once this reference image is computed, the reconstruction for the t[th] frame is solved via the following optimization problem,

$$\min_{x} \alpha \|\Psi_1(x_t - x_0)\|_p^p + (1-\alpha)\|\Psi_2 x_t\|_p^p \text{ subject to } y_t = A_t x_t \tag{19}$$

Here, $\Psi_1$ and $\Psi_2$ are sparsifying transforms (wavelet or gradient). The $l_p$-norm ($0<p\leq 1$) is the sparsity promoting objective function. Here there are two sparsity promoting terms. The first term, assumes that the difference between the current frame and the reference image is sparse in $\Psi_1$. The second term assumes that the t[th] frame is sparse in $\Psi_2$. The scalar $\alpha$ controls the relative importance of the two sparsity promoting terms.

This technique (19) is called the Prior Image Constrained Compressed Sensing (PICCS). This was originally developed with convex sparsity promoting $l_1$-norm [38, 39] but was later shown to yield even better results with non-convex $l_p$-norm (NCPICCS) [40]. It should be noted that even though the frames are reconstructed separately, this is an offline technique. This is because the reference image $x_0$ can only be generated after all the full sequence has been collected.

### 2.3. Autoencoder

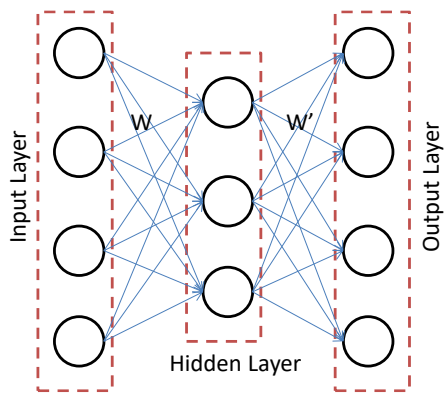

Fig. 1. Simple Single Layer Autoencoder

An autoencoder consists (as seen in Fig. 1) of two parts – the encoder maps the input to a latent space, and the decoder maps the latent representation to the data. For a given input vector (including the bias term) $x$, the latent space is expressed as:

$$h = Wx \tag{20}$$

Here the rows of W are the link weights from all the input nodes to the corresponding latent node. The mapping can be linear, but in most cases it is non-linear. Usually a sigmoid function is used, leading to:

$$h = \phi(Wx) \tag{21}$$

The sigmoid function shrinks the input (from the real space) to values between 0 and 1. Other non-linear activation functions (like tanh) can be used as well.

The decoder portion reverse maps the latent variables to the data space.

$$x = W'\phi(Wx) \tag{22}$$

Since the data space is assumed to be the space of real numbers, there is no sigmoidal function here.

During training the problem is to learn the encoding and decoding weights – W and W'. In terms of signal processing lingo, W is the analysis operator and W' is the synthesis operator. These are learnt by minimizing the Euclidean cost:

$$\arg\min_{W,W'} \|X - W'\phi(WX)\|_F^2 \tag{23}$$

Here $X = [x_1 | ... | x_N]$ consists all the training sampled stacked as columns. The problem (23) is clearly non-convex. It is solved by gradient descent techniques since the sigmoid function is smooth and continuously differentiable.

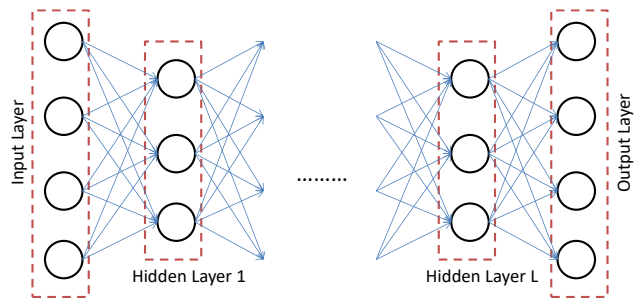

Fig. 2. Stacked Autoencoder

There are several extensions to the basic autoencoder architecture. Stacked / Deep autoencoders have multiple hidden layers (see Fig. 2). The corresponding cost function is expressed as follows:

$$\underset{W_1...W_{L-1},W'_1...W'_L}{\arg\min} \|X - g \circ \| \tag{24}$$

where $g = W_1'\phi(W_2'...W_L'(f(X)))$ and $f = \phi(W_{L-1}\phi(W_{L-2}...\phi(W_1 X)))$

Solving the complete problem (24) is computationally challenging. The weights are usually learned in a greedy fashion – one layer at a time [41, 42].

Stacked denoising autoencoders [43] are a variant of the basic autoencoder where the input consists of noisy samples and the output consists of clean samples. Here the encoder and decoder are learnt to denoise noisy input samples.

Another variation for the basic autoencoder is to regularize it, i.e.

$$\underset{(W)s}{\arg\min} \|X - g \circ \quad + R(W, X) \tag{25}$$

The regularization can be a simple Tikhonov regularization – however that is not used in practice. It can be a sparsity promoting term [44] or a weight decay term (Frobenius norm of the Jacobian) as used in the contractive autoencoder [45]. The regularization term is usually chosen so that they are differentiable and hence minimized using gradient descent techniques.

## 3. Proposed Approach

First we will discuss the non-linearity of CS inversion technique. Then we will briefly discuss the motivation behind our approach by drawing examples from MRI and CT reconstruction.

### 3.1. Non-Linearity of CS Inversion

In general CS addresses the problem of solving an under-determined linear inverse problem. To make the notations simpler, we express it as,

$$y_{m \times 1} = A_{m \times n} x_{n \times 1}, m < n \tag{26}$$

Although x is a high dimensional vector, it is supposed to be s-sparse, i.e. will have only 's' non-zero elements the rest 'n-s' being zeroes.

Ideally one would like to solve the $l_0$-minimization problem (27), but as is well known, this is NP hard.

$$\min_{x} \|x\|_0 \text{ subject to } y = Ax \tag{27}$$

One way to solve (27) is to employ greedy techniques such as orthogonal matching pursuit (OMP) [46]. This is a greedy approach which detects one support (non-zero position in x) at a time and estimates its value. The full algorithm is given by,

**Algorithm OMP**

---

**Input**: y, A, k (support)
**Initialize**: $r = y$, $\Omega = \varnothing$ (support set)
 **Repeat for k iterations**
    Compute Correlation: $c = abs(A^T r)$
    Detect Support: $l = \arg\max_{i} c_i$
    Update Support: $\Omega = \Omega \cup l$
    Estimate values at support $\Omega$: $x_\Omega = \min_{x} \|y - A_\Omega x_\Omega\|_2^2$
    Compute residual: $r = y - A_\Omega x_\Omega$
**End**

---

Here the subscript $\Omega$ means that only those columns in A indexed in $\Omega$ are selected. After the iterations are over, we get a solution with the values at some non-zero positions. To get the full x, one needs to fill in the other positions with 0 values.

OMP is a non-linear operation. In every iteration, one needs to compute the 'max' during the support detection stage – this is a highly non-linear operation. Extensions of OMP like StOMP [47], or CoSamp [48] are also non-linear. StOMP requires a thresholding operation; CoSamp requires a sorting – both are non-linear operations.

So far we have talked about greedy algorithms for sparse recovery. The more popular technique is to relax the NP hard $l_0$-norm to its closest convex surrogate the $l_1$-norm. This enjoys stronger theoretical guarantees as well. In practice the solution is obtained via (5); we repeat it for the sake of convenience in simpler notations.

$$\min_{x} \|y - Ax\|_2^2 + \lambda \|x\|_1 \tag{28}$$

Consider the simplest technique to solve (20) – Iterative Soft Thresholding Algorithm (ISTA) [49]. Every iteration (say k) consists of two steps. The first step is the Landweber Iteration (29) and the second step is the soft thresholding (30).

$$b = x_{k-1} + \sigma A^T (y - Ax_{k-1}) \qquad (29)$$

$$x_k = sign(b) \max\left(0, |b| - \frac{\lambda \sigma}{2}\right) \qquad (30)$$

where σ is inverse of the maximum Eigenvalue of $A^T A$.

The first step (21) is a simple gradient descent step – it is a linear operation. But the second step involves thresholding and is hence a non-linear operation.

To summarize, all CS recovery techniques are non-linear inversion operations.

## 3.2. MRI Reconstruction

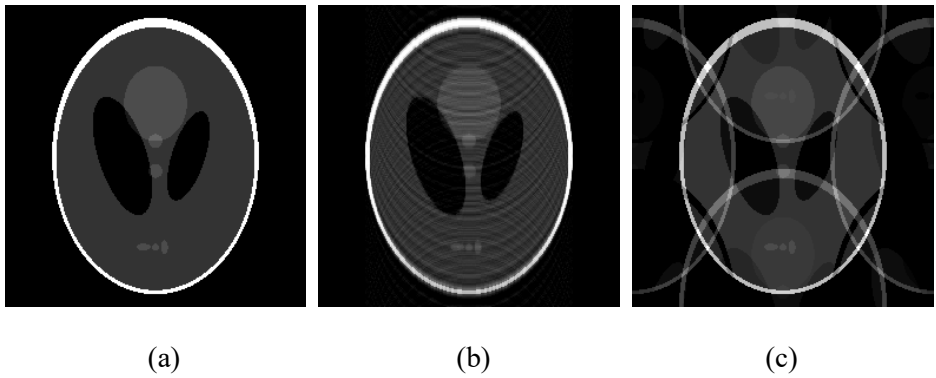

(a)　　　　　　　　(b)　　　　　　　　(c)

Fig. 3. (a) Original, (b)Reconstruction from Variable Density Random Sampling, (c) Reconstruction from Periodic Undersampling

In Fig. 3 we show what happens when the Fourier space is undersampled. When the Fourier coefficients are randomly undersampled, and the image is reconstructed by zero-filling (at missing locations) followed by an inverse FFT, we get an image like the one in Fig. 1b. The aliasing artifacts are there but they are unstructured and appear more like noise. When the Fourier coefficients are periodically undersampled and reconstructed using inverse FFT after zero-filling, we get an aliased image like the right one in Fig. 1c. This is structured aliasing.

Compressed Sensing, cannot recover the image when the K-space is periodically undersampled. Most of the theory in CS is built on the concept of randomness. With random undersampling, the aliasing artifacts are manifested like noise and CS can remove them. To quote Donoho [47] '*noiseless under-determined problems behave like noisy well determined problems*'.

Fundamentally CS acts like a denoising algorithm. This is especially true when we realise the relationship between CS recovery algorithms like ISTA and soft thresholding based denoising techniques [50]; the former applies the latter at in an iterative fashion. Theoretical studies in CS show that for most sparse signals, matrices and operators, amenable to CS recovery behave as nearly orthogonal systems; 'nearly' in the sense that it does not yield the actual sparse signal when the transpose of the operator is applied on the measured samples, but rather recovers a noisy version of the sparse signal (21). In the next step, CS algorithms denoise the signal (22) to yield a solution closer to the actual.

It has been shown in [44] that it is possible for autoencoders to learn a mapping from a denoised signal to a clean signal so that the learnt autoencoder when applied on a new noisy signal can yield a denoised representation of it. The idea of denoising autoencoders is not new, but prior studies like [43] did not use it explicitly for denoising; rather they were interested in learning robust mappings for learning representations (to be used in classification).

In this work, we propose a similar idea. Our goal is to solve (2). We repeat it for the sake of convenience:

$y = RFx$

Instead of employing a well designed non-linear inversion operation like CS, we will learn a de-aliasing autoencoder so as to remove unstructured aliasing artifacts from zero-filled reconstructed images (31).

$$\hat{x} = F^H R^T y \qquad (31)$$

These aliased images $\hat{x}'s$ are input to the autoencoder; at the output are their clean versions $x's$. Given a large number of training samples, the autoencoder learns a non-linear mapping between the input aliased samples and the clean output.

During operation / testing, a zero-filled reconstructed image will be provided at the input, and we expect the autoencoder to generate a clean version of it at the output.

### 3.3. CT Reconstruction

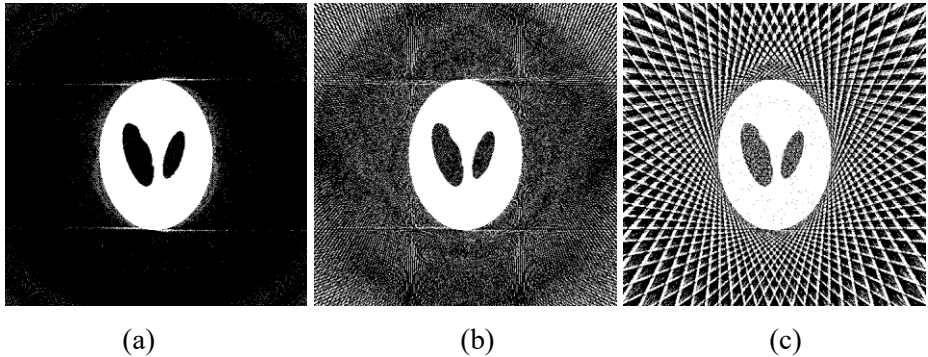

(a) (b) (c)

Fig. 4. (a) Very Dense (0.2 degrees) projections (b) Dense (1 degree) projections (c) Parsimonious (5 degrees) Sampling

Fig. 4. shows the reconstruction of an image from very dense projections (0.2 degree gap) to dense projections (1 degree gap) to parsimonious projections (5 degrees gap). The images have been reconstructed using filtered back projection (FBP). One can easily see how the reconstruction quality degrades (artifacts increase) as the angular gap between projections increase.

CS works on the CT reconstruction problem in the same manner as before. The first step is a linear gradient descent, followed by removal of artifacts via thresholding.

In this work, we treat reconstruction as a de-aliasing problem as before. From the parsimoniously sampled projections, we get an aliased image by FBP. Such aliased images form the input training data to the stacked autoencoder. During training the output is the corresponding clean image. The autoencoder will learn to remove the aliasing artifacts. During testing, when a FBP reconstructed image is input to the autoencoder, we expect to get a clean, de-aliased output.

### 3.4. Robust Autoencoding

The standard autoencoder formulation employs a Euclidean cost function. This is optimal, when the error is small / Normally distributed. Unfortunately, this is not the case for our problem – unstructured aliasing artifacts cannot be modelled as a Normal distribution. The images suggest that the error is usually large but sparse. Hence, the $l_2$-norm (Euclidean) minimization technique used in the standard autoencoder will fail, given such outliers. In statistics there is a large body of literature on robust estimation. The Huber function [51] has been in use for more than half a century in this respect. The Huber function is an approximation of the more recent absolute distance based measures ($l_1$-norm). Recent studies in robust estimation prefer minimizing the $l_1$-norm instead of the Huber function [52-54]. The $l_1$-norm does not bloat the distance between the estimate and the outliers and hence is robust. The problem with minimizing the $l_1$-norm is computational. However, over the years various techniques have been developed. The earliest known method is based on Simplex [55]; Iterative Reweighted Least Squares [56] used to be another simple yet approximate technique. Other approaches include descent based method introduced by [57] and Maximum Likelihood approach [58]. Practical studies [59, 60] have empirically shown the superiority of $l_1$-norm (over Euclidean norm) in removing outliers.

Following the studies on robust estimation, we propose a robust version of the autoencoder based on minimizing the $l_1$-norm. Our formulation is,

$$\arg\min_{W,W'} \|X - W'\phi(WX)\|_1 \tag{31}$$

This is a non-smooth function. Hence simple backpropagation type gradient based techniques will not work. Moreover, such techniques are decades old and not very efficient. In this work, we propose to solve this problem using the Split Bregman approach.

First we substitute, $P = X - W'\phi(WX)$; thus converting (31) to the following,

$$\underset{P,W,W'}{\arg\min} \|P\|_1 \text{ such that } P = X - W'\phi(WX) \tag{32}$$

The unconstrained Lagrangian is given by,

$$\underset{P,W,W'}{\arg\min} \|P\|_1 + L^T\left(P - (X - W'\phi(WX))\right) \tag{33}$$

The Lagrangian imposes equality at every step; this is too stringent a requirement in practice. One can relax the equality constraint initially and enforce it only during convergence. This is the Augmented Lagrangian formulation (34),

$$\underset{P,W,W'}{\arg\min} \|P\|_1 + \lambda \|P - (X - W'\phi(WX))\|_F^2 \tag{34}$$

In the next step, we make another substitution $Z = \phi(WX)$ and write down the Augmented Lagrangian for the same.

$$\underset{P,W,W',Z}{\arg\min} \|P\|_1 + \lambda \|P - (X - W'Z)\|_F^2 + \mu \|Z - \phi(WX)\|_F^2 \tag{35}$$

The problem with the Augmented Lagrangian approach is that, one needs to solve the full problem for every value of λ and μ; and keep on increasing them in order to enforce equality at convergence – this is time consuming. Besides, increasing the values of these hyper-parameters is heuristic. A more elegant approach is to introduce Bregman relaxation variables $B_1$ and $B_2$.

$$\underset{P,W,W',Z}{\arg\min} \|P\|_1 + \lambda \|P - (X - W'Z) - B_1\|_F^2 + \mu \|Z - \phi(WX) - B_2\|_F^2 \tag{36}$$

Although this problem is not completely separable, we can segregate (36) into alternate minimization of the following subproblems.

$$\text{P1:} \underset{P}{\arg\min} \|P\|_1 + \lambda \|P - (X - W'Z) - B_1\|_F^2$$

$$\text{P2:} \underset{W}{\arg\min} \|Z - \phi(WX) - B_2\|_F^2 \equiv \underset{W}{\arg\min} \|\phi^{-1}(Z - B_2) - WX\|_F^2$$

$$\text{P3:} \underset{W'}{\arg\min} \|P - (X - W'Z) - B_1\|_F^2$$

$$\text{P4:} \underset{Z}{\arg\min} \|Z - \phi(WX) - B_2\|_F^2$$

Subproblems P2-P4 are simple linear least squares problems. They have analytic solutions. Subproblem P1 is an $l_1$-minimization problem. This too has a closed form solution in the form of soft thresholding [50]. The last step is to update the Bregman relaxation variables:

$$B_1 \leftarrow P - (X - W'Z) - B_1 \tag{37a}$$

$$B_2 \leftarrow Z - \phi(WX) - B_2 \tag{37b}$$

The problem is non-convex thus there is no guarantee of reaching a global optimum. In this case, we continue the iterations till the objective function does not change significantly in subsequent iterations. We also have a cap on the maximum number of iterations; we have kept it to be 500.

## 4. Experimental Results

### 4.1. Experiments on Synthetic Data

First we show the denoising performance on the CIFAR-10 dataset. The 50,000 training images were used to learn the autoencoder and the remaining 10,000 test images were used for testing denoising performance. The color images were converted to grayscale and the pixel values were normalized between 0 and 1. We compare the results with the standard Denoising Autoencoder (DAE) [43] and with the sparse DAE [44]. In [44] it was shown that sparsification improves the quality of image denoising. For all the autoencoders the best results are reported. The DAE yields best results for 512 nodes, the sparse DAE yields best results for 1024 nodes and our proposed robust DAE yields best results for 786 nodes. The DAE and the sparse DAE yields best results for sigmoid activation function; our proposed robust DAE yields the best results for tanh.

We carry out experiments on two problems – denoising and de-aliasing. For the denoising problem, we test on impulse noise; where a number of pixel values in the image are corrupted with 0's or 1's. In these experiments 15% of the total pixels in the image are corrupted. For the de-aliasing problem, the Fourier frequency space is randomly under-sampled and the aliased images are produced by zero-filling; in this problem we sample randomly sample 50% of the Fourier coefficients. The gold standard for such inverse problems is compressed sensing (CS); for CS recovery it is assumed that the images are sparse in wavelet domain. For the impulse denoising problem, instead of using the standard l2-norm data fidelity as used in CS, we employ the $l_1$- $l_1$ ($l_1$-norm data fidelity and $l_1$-norm sparsity) technique [61] since it is optimum for impulse noise.

Table 1. Denoising and De-aliasing on CIFAR-10

| Problem Type | Compressed Sensing | | Proposed | | DAE | | Sparse SDAE | |
|---|---|---|---|---|---|---|---|---|
| | PSNR | SSIM | PSNR | SSIM | PSNR | SSIM | PSNR | SSIM |
| Denoising | **37.78** | **0.98** | **37.53** | **0.98** | 23.99 | 0.73 | 25.34 | 0.78 |
| De-aliasing | **27.63** | **0.81** | 23.32 | 0.76 | 20.81 | 0.60 | 21.06 | 0.66 |

The results are as expected. CS yields the best results. However for the denoising problem, our results are comparable to CS (a difference less than 0.5 dB PSNR is imperceptible). For the denoising problem, our method and CS yields far superior results compared to other autoencoders because we use the $l_1$-norm data fidelity; while other autoencoders use $l_2$-norm. – this is not optimal for impulse noise.

For the de-aliasing problem, we do better than other deep learning techniques, but is slightly worse than CS. However, it must be remembered that we are using random under-sampling of the Fourier frequency space for these experiments; this is optimal for CS. But in real life, MRI uses more structured under-sampling. As we will see in the next set of experiments, the gap between CS and our proposed method decreases for practical scenarios.

## 4.2. Dynamic MRI Reconstruction

In an unpublished work by one of the authors [61], a stacked denoising autoencoder was used to solve the same problem. Therein all the images were resized to the same size and a denoising autoencoder was learnt. The problem with this approach is that it will only work with images of a single size. For images of different size, one need to train different autoencoders. Such an architecture is not generalizable. To overcome this limitation, we move to a patch based reconstruction technique. The steps are as follows:

**Training**

- Zero fill the unsampled locations of the K-space.

- Invert the zero-filled K-space using inverse NUFFT (for non-Cartesian K-space acquisition).
- Extract non-overlapping patches of size 32x32 from these images.
- Train the autoencoder with the aliased patches at the input and corresponding clean patches at the output.

**Testing**

- Zero-fill the unsampled locations of the K-space.
- Invert the zero-filled K-space using inverse NUFFT (non-uniform fast Fourier transform).
- Take non-overlapping patches of 32x32 and input them to the trained autoencoder.
- The patches at the output need to aligned and post-processed to remove blocking artifacts.

The experimental data for training was obtained from the Laboratory of Neuro Imaging (LONI) at the University of Southern California. The dataset contains about 17424 volumes; and multiple slices in each volume. In total we have used about 500,000 patches from 100,000 images for training the autoencoders. Many of these images have been post-processed and enhanced, but that does not pose to be an issue. The goal is to learn a non-linear mapping; one could alternately use the popular image-net dataset as well. It has been seen in deep learning problems that autoencoders learnt on the image net can yield extremely good classification and denoising results on completely different image datasets. Basically the autoencoder learns an approximation map from the aliased patches to the clean patch. One may question, why we did not go deeper by using the stacked autoencoder approach. Our goal is to learn (approximately) a non-linear inversion operation. This can be done by a single layer, given enough training data. The stacked architecture is used when the training data is not enough, and there might be the issue of over-fitting. Also deeper layers are assumed to learn more abstract representations which is beneficial for classification tasks. In this case, we have enough training samples (500,000), and we do not need to learn abstract representation – our goal is to approximate a non-linear inversion operation; therefore a single layer architecture suffices. Our single layer architecture has 4096 hidden nodes.

The experimental / test data is NOT from the LONI (training) dataset; evaluation was performed on five other datasets. Two myocardial perfusion MRI datasets were obtained from [63]. We will call these two sequences to be called the Cardiac Perfusion Sequence 1 and 2. The data was collected on a 3T Siemens scanner. Radial sampling trajectory was used; 24 radial sampling lines were acquired for each time frame. The full resolution of the dynamic MR images is 128 x 128. About 6.7 samples were collected per second. The scanner parameters for the radial acquisition were TR=2.5–3.0 msec, TE=1.1 msec, flip angle = 12° and slice thickness = 6 mm. The reconstructed pixel size varied between 1.8 mm2 and 2.5 mm2. Each image was acquired in a ~ 62-msec read-out, with radial field of view (FOV) ranging from 230 to 320 mm.

The third and fourth datasets comprise of the Larynx and Cardiac sequence respectively. The data has been obtained from [64]. The larynx sequence is of size 256 x 256 and the cardiac sequence is of size 128 x 18 for each time frame. Six images were collected per second. We do not have further details pertaining to the MRI acquisition for these datasets. In this work, we have simulated radial sampling with 24 lines (for keeping it at par with the aforementioned dataset) for each image.

Our final dataset is obtained from [65]. It consists of a dynamic MRI scan of a person repeating the word 'elgar'. This is the Speech Sequence. The image is of resolution 180 x 180 and is obtained at the rate of 6 frames per second. This dataset was collected to study the tongue positions during speech. The MRI acquisition parameters related to this study are not reported here. For this sequence, we simulated radial K-space sampling with 24 lines.

The reconstruction accuracy is measured in terms of Normalised Mean Squared Error (NMSE). The proposed technique is compared with one offline and one online technique. The offline technique is k-t SLR [21] – this has been the de facto gold standard for dynamic MRI reconstruction. The online techniques we compared against is Differential CS [31]; it has been shown in [31] that the method proposed therein is faster and more accurate than all Kalman filtering based techniques, therefore we do not compare against such filtering based dynamic MRI reconstruction methods. We do not compare our method with the previous unpublished work [62], since it cannot handle images of different size. Also we do not compare our method with $l_2$-norm autoencoders such as denoising autoencoder or sparse denoising autoencoder; since in the previous sub-section we have already shown that our proposed RObust DE-aliasing autOencoder (RODEO) yields better de-alisaing results than these. For removing blocking artifacts we use the algorithm proposed in [66]; the implementation for the same is available in [67]. The reconstruction errors and the standard deviations are shown in the following Table.

Table 1. Mean and the Standard Deviation of the NMSEs on different datasets

| Dataset | k-t SLR [21] | Diff CS [31] | RODEO |
|---|---|---|---|
| Cardiac Perfusion 1 | 0.120, ±0.040 | 0.175, ±0.067 | 0.172, ±0.045 |
| Cardiac Perfusion 2 | 0.181, ±0.061 | 0.221, ±0.089 | 0.202, ±0.052 |
| Larynx | 0.024, ±0.012 | 0.030, ±0.015 | 0.031, ±0.010 |
| Cardiac | 0.073, ±0.022 | 0.076, ±0.028 | 0.075, ±0.021 |
| Speech | 0.217, ±0.055 | 0.311, ±0.087 | 0.308, ±0.052 |

We observe that k-t SLR yields the best results. This is understandable - it is an offline technique and has access to the whole dataset for reconstruction. It exploits the spatio-temporal correlation from all the frames and therefore yields the best result. What is surprising is that our proposed method yields results at par with state-of-the-art CS based online reconstruction techniques like [31]. In fact we will show in the next table that our results are actually better. NMSE has been widely used for evaluating CS reconstruction; but it is not the best indicator of visual quality. Structural Similarity index (SSIM) [67] is a metric appropriate choice for evaluating visual quality. In the next table, we show the SSIM results. We find that k-t SLR yields the best SSIM – this is expected. Our proposed method yields better results than the previous CS based online reconstruction technique.

Table 2. Mean SSIMs on different datasets

| Dataset | k-t SLR [21] | Diff CS [31] | RODEO |
|---|---|---|---|
| Cardiac Perfusion 1 | 0.85 | 0.78 | 0.79 |
| Cardiac Perfusion 2 | 0.87 | 0.80 | 0.82 |
| Larynx | 0.93 | 0.85 | 0.88 |
| Cardiac | 0.94 | 0.86 | 0.88 |
| Speech | 0.76 | 0.70 | 0.72 |

For visual evaluation the difference images are shown. This is the standard method to visually evaluate image reconstruction. These are shown in Fig. 5. The contrast is magnified 10 times for visual evaluation. Visual evaluation corroborates the numerical results. The k-t SLR method yields the best reconstruction – the difference images are the darkest. Our method yields better results than the differential CS method.

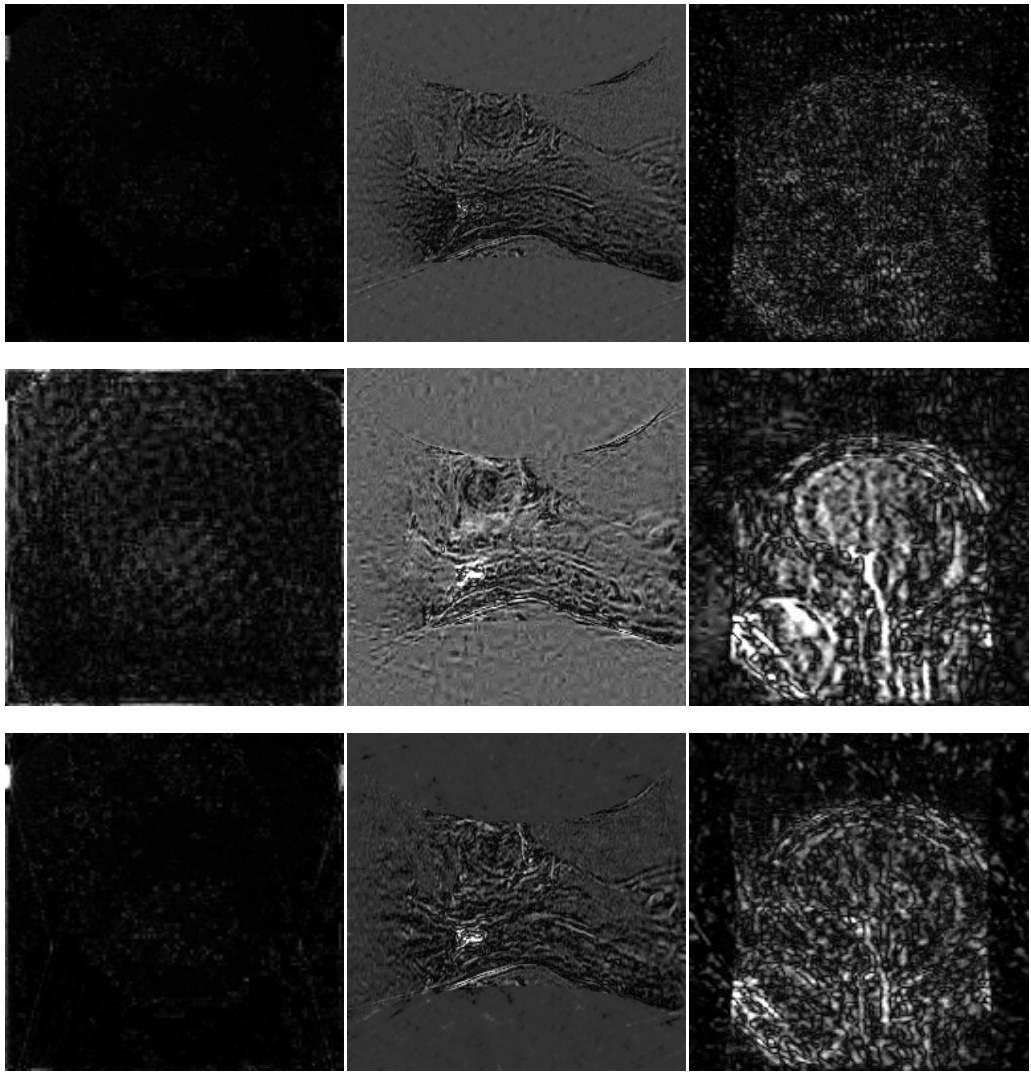

Fig. 5. Difference Images. Left to Right - Cardiac, Larynx and Speech. Top to Bottom - k-t SLR; Differential CS Reconstruction; Robust DAE Reconstruction

The experiments were run on a Core i7 CPU (3.1 GHz) having a 16 GB RAM running 64 Windows 7. The platform is Matlab 2012a with parallel processing enabled. We ran four parallel processes, i.e. for images of size 128x128, 16 non-overlapping patches of size 32x32. De-aliasing each patch takes about 0.006 seconds on an average; 4 such patches are processed simultaneously. The total time taken to process the entire image is 0.027 seconds. Differential CS [31] takes about 0.15 seconds to process a frame of same size.

For images of size around 128x128 pixels, the data acquisition rate is usually between 6 to 7 frames per second (at full resolution). With acceleration (partial K-space sampling) this frame rate can go up higher - 18 to 30 frames per second for 15 to 24 radial sampling trajectories. Differential CS can only reconstruct 6-7 frames per second; this is three times slower than the acquisition rate. Our method can recover the more than 30 frames per second; this is much beyond the data acquisition speed.

### 4.3. CT Reconstruction

For CT, we do not have access to dynamic datasets. Therefore we experiment on a standard static dataset. The point we need to establish – speedy reconstruction with minimal loss in quality; can be thoroughly established using the static dataset.

We use the same LONI dataset to train. The training and the testing procedures remain almost the same; the only change is in the first step of training and testing. We assume parallel beam tomography. Therefore to synthesize the projections, we use the simple radon transform. For inversion from parsimonious CT projections filtered back projection (FBP) is employed. This yields an image with spoke-like aliasing artifacts. Other analytic techniques have evolved for CT reconstruction (from radon transform) [67, 68], but for our problem the fast and simple FBP suffices. From such a reconstruction, non-overlapping patches of 32x32 are extracted and used as training inputs to the autoencoder. At the output are the corresponding clean patches. As before, we use 4096 nodes in the hidden layer.

The image datasets used (for testing) in this experiment were from the Laboratory of Human Anatomy and Embryology, University of Brussels (ULB), Belgium. The details of the acquisition and the data structure can be obtained from [70]. All images are of size 512x512. In total there are about 1001 images. The organization of the dataset is succinctly given in Table 3.

Table 3. Organization of CT Dataset

| Body part | Bone | Structure | # Slices | Acquisition time |
|---|---|---|---|---|
| Lower limb | Humeral bone | Proximal epiphysis | 102 | 35 |
| | | Diaphysis | 199 | 55 |
| | | Distal epiphysis | 120 | 41 |
| | Scapula | Scapula (parts 1 & 2) | 80+78=158 | - |
| | Clavicle | Clavicle (parts 1 & 2) | 80+70=150 | 42 |
| | Finger | First phalanx | 87 | 30 |
| | | Second phalanx | 63 | 22 |
| | | Third phalanx | 42 | 15 |
| Lower limb | Iliac | Crest, Ilium | 90 | 61 |
| | | Acetabulum | 132 | 45 |
| | | Ischium, pubis | 38 | 26 |
| Spine | C1 | C1 | 47 | 38 |
| | C2 | C2 | 73 | 56 |
| | C7 | C7 | 62 | 60 |
| | L3 | L3 | 95 | 37 |
| | Sacral bone | Sacral bone | 138 | 31 |

Since this is a static CT reconstruction problem, the CS based technique we compare with is [35]. The $l_1$-minimization problem is solved using the spectral projected gradient l1 algorithm [71]. We show reconstruction for two scenarios – 2.5º and 5 º difference between successive tomographic projections. The first one is a dense sampling (144 spokes); the second one is moderately sparse (72 spokes). The reconstruction results are shown in the following table.

Table 4. Reconstruction Accuracy in NMSE

| Structure | CS [35] | | Proposed RODEO | |
|---|---|---|---|---|
| | 2.5º | 5º | 2.5º | 5º |
| Proximal epiphysis | .06, ±.03 | .11, ±.09 | .11, ±.04 | .13, ±.07 |
| Diaphysis | .07, ±.03 | .15, ±.08 | .11, ±.03 | .16, ±.06 |
| Distal epiphysis | .04, ±.02 | .10, ±.06 | .09, ±.03 | .13, ±.06 |
| Scapula (parts 1 & 2) | .11, ±.09 | .18, ±.17 | .16, ±.07 | .19, ±.09 |
| Clavicle (parts 1 & 2) | .08, ±.06 | .13, ±.09 | .12, ±.05 | .15, ±.06 |
| First phalanx | .10, ±.08 | .18, ±.18 | .16, ±.08 | .21, ±.09 |

| Second phalanx | .10, ±.06 | .17, ±.12 | .17, ±.08 | .20, ±.10 |
|---|---|---|---|---|
| Third phalanx | .11, ±.10 | .19, ±.17 | .17, ±.09 | .24, ±.09 |
| Crest, Ilium | .09, ±.07 | .15, ±.09 | .15, ±.07 | .17, ±.08 |
| Acetabulum | .09, ±.02 | .16, ±.09 | .14, ±.06 | .18, ±.09 |
| Ischium, pubis | .11, ±.10 | .18, ±.15 | .16, ±.08 | .19, ±.10 |
| C1 | .07, ±.05 | .13, ±.09 | .11, ±.07 | .13, ±.09 |
| C2 | .10, ±.09 | .18, ±.16 | .16, ±.08 | .18, ±.10 |
| C7 | .11, ±.06 | .17, ±.12 | .15, ±.06 | .17, ±.08 |
| L3 | .11, ±.10 | .19, ±.17 | .17, ±.08 | .19, ±.09 |
| Sacral bone | .10, ±.10 | .19, ±.18 | .15, ±.09 | .19, ±.10 |

The results show that our propose method yields slightly worse reconstruction compared to the CS technique in terms of NMSE. The difference between RODEO and CS is larger for finer sampling (2.5º), but the difference decreases for coarser sampling (5º). However, our results are more robust; the standard deviations from our method are much smaller than those of CS. This implies that our method is more agnostic to the structure under study compared to CS.

In the following table we show the reconstruction accuracy in terms of SSIM [72]. We find that the difference between CS reconstruction and our proposed reconstruction is much smaller in terms of SSIM compared to NMSE. The same phenomenon was observed for MRI reconstruction. The reason is easy to comprehend. NMSE is based on 'mean squared error'. CS uses an $l_2$-norm cost function; so it always gives good results in terms of matrices computed on least squares. Our method minimizes the $l_1$-norm; therefore our results on mean squared based error metrics are not as good. However our method is better in terms of visual quality, we are closer to CS reconstruction than implied by NMSE values – especially at coarser sampling (5º).

Table 5. Reconstruction Accuracy in terms of mean SSIM.

| Structure | CS [35] | | Proposed RODEO | |
|---|---|---|---|---|
| | 2.5º | 5º | 2.5º | 5º |
| Proximal epiphysis | 0.93 | 0.87 | 0.88 | 0.85 |
| Diaphysis | 0.92 | 0.87 | 0.88 | 0.86 |
| Distal epiphysis | 0.94 | 0.88 | 0.87 | 0.86 |
| Scapula (parts 1 & 2) | 0.88 | 0.85 | 0.85 | 0.83 |
| Clavicle (parts 1 & 2) | 0.91 | 0.86 | 0.85 | 0.84 |
| First phalanx | 0.90 | 0.86 | 0.86 | 0.84 |
| Second phalanx | 0.90 | 0.86 | 0.87 | 0.83 |
| Third phalanx | 0.88 | 0.85 | 0.86 | 0.83 |
| Crest, Ilium | 0.89 | 0.85 | 0.86 | 0.85 |
| Acetabulum | 0.88 | 0.85 | 0.84 | 0.84 |
| Ischium, pubis | 0.88 | 0.85 | 0.85 | 0.85 |
| C1 | 0.93 | 0.88 | 0.87 | 0.85 |
| C2 | 0.91 | 0.87 | 0.87 | 0.84 |
| C7 | 0.90 | 0.86 | 0.85 | 0.85 |
| L3 | 0.89 | 0.86 | 0.86 | 0.85 |
| Sacral bone | 0.90 | 0.87 | 0.86 | 0.84 |

For visual evaluation two sample slices are shown in the following figure. These are the difference images and have been contrast enhanced for visual clarity. This pertains to 5º difference in successive projections. One can see that there is not much difference between the difference images from these two techniques.

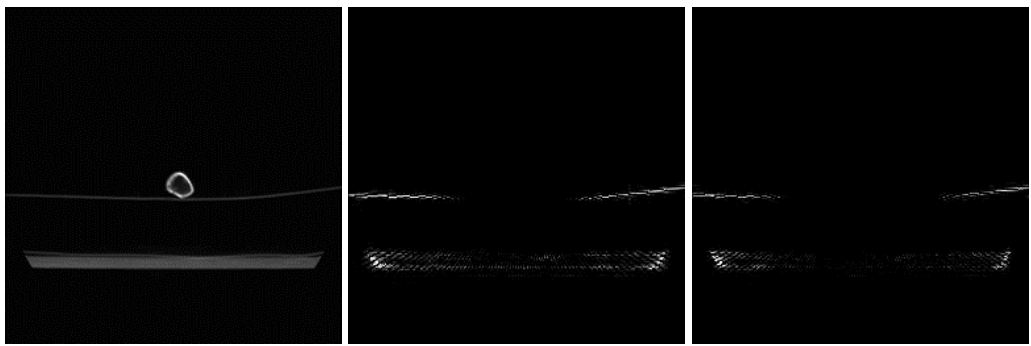

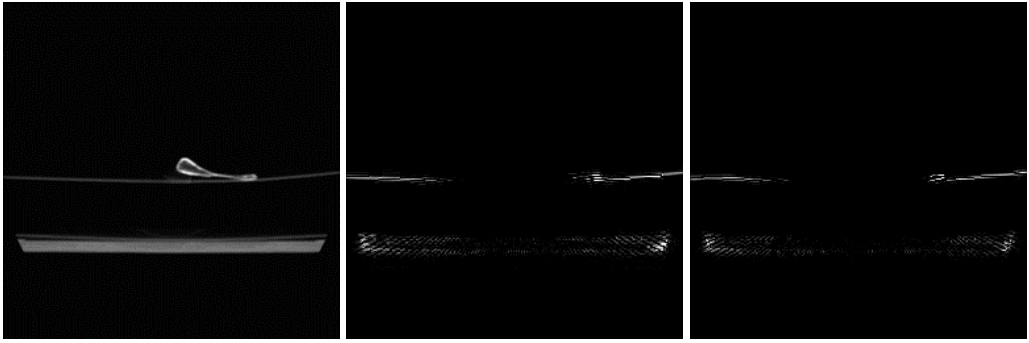

Fig. 5. Left to Right – Groundtruth, Proposed difference image and CS difference image. Top – Clavicle; Bottom - Scapula

To study the reconstruction speed, let us take one example – L3 spine. This structure has 95 slices and the acquisition time is 37 seconds. The full 3D CS reconstruction time is 314 seconds. Our method takes 41 seconds for the full volume. For another example – clavicle having 150 slices in all, CS reconstruction takes 427 seconds. Our method takes 64 seconds; the acquisition time is 42 seconds. Similar improvement in reconstruction speed can be seen for the other portions of the dataset as well. Although our proposed method is not exactly real-time, we believe that the speed can be further improved by moving to a more efficient environment like C/C++ and using GPUs for matrix-vector products.

## 5. Conclusion

This is the first work that proposes MRI and CT reconstruction using autoencoder. Usually autoencoders are used to pre-train deep neural networks. There are some studies on the denoising (Gaussian noise) capacity of autoencoders. Extending on the simple denoising framework, this paper solves a complex inverse problem. Instead of formulating the inversion operation (as an optimization problem) it 'learns' the inversion from examples.

More specifically we are interested reconstructing MR images from sub-sampled K-space and reconstructing CT images from parsimonious tomographic projections. This is an under-determined inverse problem. Compressed Sensing (CS) is usually used to solve such problems. This work proposes an alternate approach. A crude inversion is obtained (from the K-space / tomographic projection) by an analytic technique (inverse FFT or FBP); the inversion is crude in the sense, it has a lot of reconstruction artifacts arising from aliasing. An autoencoder learns to map the aliased image to a clean image using a large number of training samples. Once the learning is complete, a test image (obtained by the crude inversion) when input to the de-aliasing autoencoder, produces a clean version of it.

Our approach stems from past theoretical studies by Kolmogorov [2], Kukrova [3], Hornik [4] and Cybenko [5]. These studies showed that given enough training samples, neural networks can learn arbitrary non-linear function maps. In this case we are interested in learning the mapping from an aliased image to a clean image.

However, the basic formulation for autoencoder is not suitable for our problem. It uses a Euclidean cost function – this is optimal when the deviations are small / Normally distributed. Aliasing does not follow such a distribution; it appears mostly as outliers (large but sparse). Therefore we learn an autoencoder using the more robust $l_1$-norm cost function instead. The resulting problem is non-smooth hence standard gradient / back-propagation based techniques cannot be used here. We solve the ensuing problem using the Split Bregman technique.

Experiments are carried out on real MRI and CT reconstruction problems. Comparison is done with CS based techniques. We find that our proposed method yields slightly lower reconstruction quality compared to CS but is significantly faster. This is because CS requires solving an iterative optimization problem; our proposed technique requires only a few matric matrix vector multiplications. We show that our method is fast enough for real-time dynamic MRI reconstruction – we can reconstruct more than 30 frames per second; this is much beyond the acquisition capacity of current MRI scanners and beyond the reach for CS based reconstruction. For CT reconstruction, the acquisition time is faster than our reconstruction speed (CT is few orders of magnitude faster than MRI in terms of acquisition); but we believe that moving to a more efficient platform (current experiments are reported on Matlab) and utilizing the computational resources of GPU, we can accelerate CT reconstruction even further and pave way for real-time imaging.